\documentclass[12pt]{article}

\usepackage{amsmath}
\usepackage{amscd}
\usepackage{amsfonts}
\usepackage{amssymb}
\usepackage{amsthm}
\usepackage{epsfig}

\renewcommand {\d}  {\partial}
\newcommand {\vd}   {\delta}
\renewcommand {\phi}{\varphi}

\newcommand {\s}  {\sigma}

\newcommand {\e}   {\epsilon}

\newcommand {\lb} {\left (}
\newcommand {\rb} {\right )}

\newcommand{\la}{\langle}
\newcommand{\ra}{\rangle}

\renewcommand{\leq}{\leqslant}

\newcommand{\f}{\frac}

\newcommand{\pexp}{\mathop{\rm P\,exp}}

\newcommand{\dt}[1]{\dot{#1}}
\newcommand{\dtt}[1]{\ddot{#1}}
\newcommand{\ds}[1]{{#1'}}
\newcommand{\dss}[1]{{#1''}}
\newcommand{\dst}[1]{{#1'''}}
\newcommand{\dsf}[1]{{#1''''}}

\newcommand{\ti}{\tilde}

\newcommand{\beq}{\begin{equation}}
\newcommand{\eeq}{\end{equation}}

\newcommand{\bal}{\begin{aligned}}
\newcommand{\eal}{\end{aligned}}

\newcommand\bqa {\begin{eqnarray}}
\newcommand\eqa {\end{eqnarray}}

\def\appendix#1{
  \addtocounter{section}{1}
 \setcounter{equation}{0}
  \renewcommand{\thesection}{\Alph{section}}
 \section*{Appendix \thesection\protect\indent \parbox[t]{11.715cm} {#1}}
  \addcontentsline{toc}{section}{Appendix \thesection\ \ \ #1}
  }

\newcommand{\newsection}{
\setcounter{equation}{0}
\section}

\newcommand{\eq}[1]{\begin{equation} #1 \end{equation}}
\newcommand{\ar}[1]{\begin{eqnarray} #1 \end{eqnarray}}
\newcommand{\tr}{\mathop{\mathrm{tr}}\nolimits}

\def\e{{\,\rm e}\,}
\def\d{\partial}

\newcommand{\br}[1]{\left( #1 \right)}
\newcommand{\vev}[1]{\left\langle #1 \right\rangle}
\newcommand{\rf}[1]{(\ref{#1})}
\newcommand{\non}{\nonumber \\*}
\def\ll2{\ll}

\def\N{${\cal N}=4$ }
\def\ads{$AdS_5\times S^5$ }
\def\oj{{\cal O}_J}

\def\cor{\left\langle  W(C)  {\cal O}_J (x)\right\rangle}
\def\ep{\varepsilon}
\newcommand{\on}[2]{#1_{#2}}
\def\rr{\e^{2\on{Q}{0}}}

\title{
\hfill{\small UUITP-19/02}\\
\hfill{\small ITEP-TH-60/02}
\\~\\
Comparing strings in \ads to planar diagrams:
\\[0.3cm]
 an example }

\author{V. Pestun$^1$\thanks{\tt pestun@gate.itep.ru}~
and K. Zarembo$^2$\thanks{{\tt Konstantin.Zarembo@teorfys.uu.se}. Also at
ITEP, B.~Cheremushkinskaya 25, 117259 Moscow, Russia}
\\~~\\
$^1${\it Institute of Theoretical and Experimental Physics} \\ {\it
B.~Cheremushkinskaya 25, 117259 Moscow, Russia}
\\~~\\
$^2${\it Department of Theoretical Physics,
Uppsala University}\\ {\it
Box 803, SE-751 08 Uppsala, Sweden }
}

\begin{document}           

\maketitle

\abstract{The correlator of a Wilson loop with a local operator
in \N SYM theory can be represented by a string amplitude in
\ads. This amplitude describes an overlap of the boundary state,
which is
associated with the loop, with the string mode, which is dual to the
local operator. For chiral primary operators with a large R charge,
the amplitude can be calculated by semiclassical techniques.
We compare the semiclassical string amplitude to the SYM
perturbation theory and find an exact agreement to the first two
non-vanishing orders.
}

\newsection{Introduction}

The AdS/CFT correspondence establishes an equivalence of the
superstring theory in the \ads background and \N supersymmetric
Yang-Mills theory (SYM) \cite{review}. This equivalence is a true
duality: the interaction strength in the string sigma-model is the
inverse of the 't~Hooft coupling in the field theory, which makes
any direct comparison of strings with the field theory extremely
difficult. On the other hand, planar diagrams of the large-$N$
perturbation theory resemble discretized string world sheets
\cite{'tHooft:1973jz}, and one may think that a more direct
relationship between them exists beyond the fact that they
describe one and the same theory in different regions of parameter
space. In principle, one has to sum all planar diagrams (solve
large-$N$ SYM) in order to reach weakly-coupled stringy regime. As
became clear recently, it is still possible to make a comparison
to string theory without summing all diagrams. This can be done by
considering special states with large quantum numbers. In string
theory, such states are semiclassical without any reference to the
weakness of  interactions in the sigma-model. These states,
therefore, have a simple string description all the way down to
weak coupling where perturbative SYM theory becomes accurate.
Though this logic involves a certain stretch in the arguments, it
has worked in a number of examples \cite{BMN,GKP}. With the help
of the semiclassical string picture, a remarkable progress has
been made in understanding operators with large R charge (and also
with large spin) in the SYM theory \cite{BMN,GKP}. Correlation
functions of these operators have a beautiful world-sheet
description, and vice versa, the world-sheet dynamics of the
string states with large angular momentum on $S^5$ is encoded in
certain set of Feynman diagrams which have relatively simple
structure
\cite{BMN,Kristjansen:2002bb,Gross:2002su,Constable:2002hw,Santambrogio:2002sb}.

A local operator in the SYM theory is dual to a closed string state in \ads.
The string theory in
AdS  also has an open string sector which is associated with
Wilson loops \cite{MAL,Rey,SZ}.
By probing Wilson loops with operators that have parametrically
large R charge it is possible to reach the semiclassical regime in the open-string sector
too \cite{Zarembo:2002ph}.
Whether the semiclassical description in the open string sector captures
perturbative regime in the field theory
or still requires the 't~Hooft coupling to be large is not entirely clear. We
will address this question by comparing string-theory calculations with
lowest-order Feynman diagrams.

We consider a two-point correlator $\vev{W(C)\oj}$
of the Wilson loop with the   
operator that carries charge $J$ under a $U(1)$ subgroup of the
$SO(6)$ R-symmetry group. The operator of interest is chiral
primary \beq \oj = \f {(2\pi)^J} {\sqrt{J} \lambda^{J/2}} \tr Z^{J},
\eeq where  $Z=\Phi_1 + i \Phi_2$ is a combination of two of the
six adjoint scalars in the $SU(N)$ SYM theory. The correlator
$\vev{W(C)\oj}$ measures the weight with which operator $\oj$
appears in the local operator expansion of the Wilson loop. We
should compare this to an overlap of the closed string boundary
state created by the Wilson loop with the supergravity state dual
to the operator $\oj$ \cite{Berenstein:1999ij}. There is one
special case in which the exact answer is know: It is possible to
compute the correlator exactly for the circular Wilson loop
\cite{Semenoff:2001xp}, which is a chiral operator in a certain
sense \cite{Bianchi:2002gz,Erickson:1999uc,Drukker:2000rr}.
Because of the supersymmetry cancellations, only diagrams without
internal vertices contribute to its expectation value
\cite{Erickson:2000af,Drukker:2000rr}, as well as to its expansion
coefficients in chiral primaries \cite{Semenoff:2001xp}. These
diagrams can be explicitly resummed which yields the exact results
valid at any 't~Hooft coupling. The semiclassical string
calculations can also be done to all orders in the sigma-model
perturbation theory because of geometric symmetries of the circle
 \cite{Zarembo:2002ph}.
We consider arbitrary contours in this paper, for which the equations
of motion in the sigma-model can be solved order by order in $\lambda/J^2$,
where $\lambda$ is the string tension squared\footnote{We express the string tension 
and all other dimensionful
quantities in the units of the AdS radius.}, which according
to the AdS/CFT dictionary
 coincides with the 't~Hooft coupling in the SYM theory:
 $\lambda=g^2_{SYM}N$.
We will compare the expansion of the string amplitude in $\lambda/J^2$
to the ordinary
planar perturbation theory.


The paper is organized as follows. In sec.~\ref{not}, we set up the notations
and review how the correlator of a Wilson loop with the local operator
is computed in string theory. In sec.~\ref{SYMside}, we do a one-loop
calculation on the SYM side. The classical
solution of the sigma-model, which describes the correlator in string theory,
 is constructed in sec.~\ref{str}. We then compare the string amplitude
determined by this solution with one-loop SYM perturbation theory.
We also show in sec.~\ref{exp} that diagrams
of the SYM perturbation theory  exponentiate in accord with
predictions of the string theory. We
draw the conclusions and discuss the results in sec.~\ref{last}.

\newsection{Wilson loop correlator in string theory}
\label{not}

The supersymmetric Wilson loop operator is defined as \cite{MAL}
\beq W(C) =\tr \pexp \left[ \int ds \left(i A_{\mu} \ds{x}{}^{\mu}
+ \Phi_1 \,|\ds{x}|\right) \right]. \eeq This operator is a hybrid
of an ordinary non-Abelian phase factor and the unique scalar loop
operator which is conformally and Lorentz covariant
\cite{Makeenko:hm}\footnote{The most general Wilson loop depends
on scalars through an arbitrary linear combination
$\Phi_i\theta^i$, where $\theta^i$ is a unit six-vector, which may
depend on $s$ and thus parameterizes a closed contour in $S^5$. We
choose a particular $\theta^i$ for definiteness; results will not
be much different for any other constant $\theta^i$. It would be
interesting to consider varying $\theta^i$
\cite{MAL,Zarembo:2002an,Tseytlin:2002tr}, but this goes beyond
the scope of the present paper.}.

The Wilson loop operator is dual to a macroscopic string in \ads. The local
operator is dual to a supergravity mode. If we want to compute their correlator,
we should find vertex operator associated with the supergravity mode,
insert it into the world sheet of the string, combine it
with the propagator of the supergravity mode, and then integrate over all
string world-sheets and all
positions of the vertex operator \cite{Berenstein:1999ij}.
 In the limit when the distance between the loop
and the insertion of the local operator goes to infinity the propagator factorizes, and
we are left with the one-point correlation function in the sigma-model:
\beq\label{wo}
\la W(C) O_J(x)  \ra =\frac{1}{|x|^{2J}}
 \int d^2 \s_0 \int D{\cal X} \e^{-S_{\sigma m}[{\cal X} ]} V_J({\cal X} (\s_0))
+o(1/|x|^{2J}),
\eeq
where ${\cal X} $ denotes coordinates, bosonic and fermionic, of the
string in the
AdS superspace; $S_{\sigma m}[{\cal X} ]$ is the Green-Schwarz action of the
AdS sigma-model; and
$V_J({\cal X} (\s))$ is the vertex operator which creates string mode
dual to the operator $\oj$. We assume that the conformal gauge is already
fixed and regard corresponding
Faddeev-Popov factor as a part of the measure in the path integral.

The metric of $AdS_5$ in the P\'oincare coordinates is
\eq{
ds^2=dp^2+\e^{2p}dx^2.
}
 The boundary is at $p=\infty$
and the horizon of $AdS_5$ is at $p=-\infty$.
We adopt the following parameterization
of $S^5$: $\theta^i=(\cos\psi\,\cos\varphi,\cos\psi\,\sin\varphi,\sin\psi\,{\bf n})$,
where ${\bf n}$ is a unit four-vector. The sigma-model action in this
parameterization is
\eq{\label{smaction}
S_{\sigma m}=\frac{\sqrt{\lambda}}{4\pi}\int d^2\sigma\left[
\br{\d P}^2+\e^{2P}\br{\d X^\mu}^2+\br{\d\varPsi}^2+\cos^2\varPsi\br{\d\varPhi}^2
+\ldots
\right].}
The action for the rest of the angles in $S^5$ and for the world-sheet fermions
will not be important for us.
The operator $\oj$ is chiral and hence corresponds to the supergravity
(ten-dimensional massless) mode. The vertex operator of such string state
is a solution of the massless wave equation in \ads. Symmetries and the form of the
dual SYM operator $\oj$ essentially determine the necessary solution:
 it should have zero momentum
along the boundary, since we consider the large-distance
asymptotic of the correlator;  it should scale as
$V\rightarrow\e^{-J\omega}V$ under $P\rightarrow P+\omega$, since
the operator $\oj$ has dimension $J$; and it should transform as
$V\rightarrow\e^{iJ\varphi}V$ under an $SO(6)$ rotation
$\varPhi\rightarrow\varPhi+\varphi$.
These simple arguments
 determine the vertex operator
with an exponential accuracy, which will be sufficient for
our purposes:
\eq{\label{vexp}
V\propto
\e^{iJ\varPhi-JP}.
}
Fixing normalization is possible \cite{Berenstein:1999ij},
but requires much more work.

We will calculate the correlator \rf{wo} in the double-scaling
limit of large string tension and large R charge with their ratio
$j=J/\sqrt{\lambda}$ fixed. The string path integral is semiclassical
in this limit, so we need just to solve classical equations of
motion in the sigma-model, which fortunately leaves behind many
hard questions like correct normalization of the measure in the
path integral or
 exact form of the vertex operator. Since the exponent in \rf{vexp}
is of the same order as the action, we should treat the action and
the vertex operator on the same footing \cite{strings2002}.
This effectively adds
a source to the action:
$$
S_{\rm eff}=S_{\sigma m}+J\br{P(\s_0)-i\varPhi(\s_0)}\equiv JS.
$$
The action with the source term added is the functional that we should
minimize in order to compute the string amplitude in the semiclassical
approximation.
The action evaluated on the classical solution
determines the correlator
$\la W(C) O_J(x)  \ra$ with an exponential accuracy at
large $\lambda$ and large $J$. Since the rescaled
action $S$ and, consequently,  the equations of motion depend on $\lambda$
and $J$ only through the ratio $j=J/\sqrt{\lambda}$, the
correlator will have the form:
\eq{\label{gstring}
\la W(C) O_J(x)  \ra\simeq\frac{1}{|x|^{2J}}\e^{-JS(j;C)}.
}

\begin{figure}[h]
\begin{center}
\epsfxsize=3cm
\epsfbox{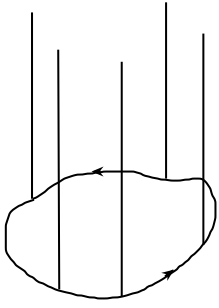}
\end{center}
\caption[x]{\small Correlator $\la W(C) O_J(x)  \ra$
to the leading order of perturbation theory.}
\label{pert0}
\end{figure}

Solving classical equations of motion of the AdS sigma-model for
generic boundary conditions is a hard problem, but it is possible
to find an approximate solution when the parameter $j$ is large.
Then the minimal surface is a cylinder to a first approximation
\cite{Zarembo:2002ph}: the shape of the contour is the same in any
slice of AdS parallel to the boundary. It was shown in
\cite{Zarembo:2002ph} that this solution predicts the following
dependence of the correlator on $J$ and $\lambda$:
\eq{\label{lopt} \la W(C) O_J(x)
\ra\sim\frac{\lambda^{J/2}}{J!}\,. } Surprisingly, this is the
same as one would get by counting combinatorial factors in the
lowest-order
 diagram (fig.~\ref{pert0}) of the SYM perturbation
theory. We should stress that the perturbative
and the string calculations are valid in different regions of parameters:
the semiclassical approximation in string theory requires both $J$ {\it and}
 $\lambda$ to be large, but the ratio $\lambda/J^2$ can be small.
The perturbation theory, strictly speaking,
 is valid only when the coupling is small. Of course, it may turn out that
 $\lambda/J^2$ is the true parameter
of perturbative expansion,
then $\lambda$ should not necessarily be small
for the perturbation theory to work.

We will develop a systematic
method to solve the sigma-model equations of motion order by order
in $1/j^2$. This will yield an expansion of the Wilson loop correlator
in $\lambda/J^2$, which resembles an ordinary perturbative
series. To compare the two, we will first compute the
one-loop correction on the SYM side.

\section{One-loop calculation in the SYM theory}\label{SYMside}

The one-loop planar correction to the correlator of a Wilson loop
with the chiral primary operator is described by the single
diagram in fig.~\ref{pert1}. The diagrams with corrections to the
external lines exactly cancel, as shown in appendix~B. Combining
the tree-level diagram \ref{pert0} with the one-loop correction
gives\footnote{Our notations and conventions for the SYM
perturbation theory are collected in appendix~A.}:
\ar{\label{intq}
\cor&=&\frac{\lambda^{J/2}}{\sqrt{J}(4\pi|x|^2)^J} \int ds_1\ldots
ds_J\,\theta_c(s_1,\ldots,s_J) \non &&\times \br{1+\lambda
\sum_{n=1}^{J}\int_{s_n}^{s_{n+1}}ds \int_s^{s_{n+1}} dr\,
G(s,r)+O(\lambda^2) } \non &=&
\frac{\sqrt{J}}{J!}\,\br{\frac{\sqrt{\lambda}}{4\pi |x|^2}}^J \non
&&\times \left[(2\pi l)^J+\lambda \int_0^{2\pi l} ds\,
\int_0^{2\pi l} dq\,(2\pi l-q)^J G(s,s+q)
 +O(\lambda^2)\right],
}
where $\theta_c$ equals one, if its arguments are cyclically ordered
along the contour, and equals zero otherwise,
 $2\pi l=\int ds\,|\ds{x}|$ is the length of the contour $C$, and
$G(s,r)$ is the sum of the gauge-boson and the scalar propagators
inserted between points $x(s)$ and $x(r)$ on the contour:
\eq{\label{prop}
G(s,r)=\frac{1-\ds{x}(s)\cdot \ds{x}(r)}{8\pi^2|x(s)-x(r)|^2}\,.
}
From now on, we use the natural parameterization of the contour $C$:
$\ds{x}{}^2=1$. In going from the second to the third line in \rf{intq},
we used the formula
\eq{\label{s!!}
\int_0^{s}ds_1\ldots\int_0^{s_{J-1}}ds_J=\frac{s^J}{J!}
}
to integrate over the end-points of external legs in the diagrams
\ref{pert0}, \ref{pert1}.

\begin{figure}[h]
\begin{center}
\epsfxsize=3cm
\epsfbox{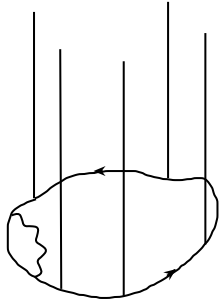}
\end{center}
\caption[x]{\small One-loop correction to the correlator $\la W(C) O_J(x)  \ra$.
The wavy line denotes the sum of the gauge-boson and the scalar propagators.}
\label{pert1}
\end{figure}

The answer considerably simplifies in the limit of large $J$, because
then
\eq{\label{f0}
\int_0^{2\pi l} dq\,(2\pi l-q)^J F(q)=\frac{(2\pi l)^J}{J}\,F(0)+O(1/J^2)
}
for any smooth function $F(q)$, and the propagator \rf{prop}
can be expanded as
\eq{
G(s,s+q)=-\frac{\dst{x}\cdot\ds{x}}{16\pi^2}+O(q)
=\frac{\dss{x}{}^2}{16\pi^2}+O(q).
}
We get:
\eq{\label{one-loop}
\cor=\frac{\sqrt{J}}{J!}\,\br{\frac{\sqrt{\lambda}\,l}{2 |x|^2}}^J
\left[1+\frac{\lambda l}{8\pi J}\int_0^{2\pi l} ds\,\dss{x}{}^2+O\br{\frac{\lambda}{J^3}}
+O(\lambda^2) \right].
}

The fact that the large-$J$ limit is only sensitive to the local
limit of the propagator has a simple explanation: since the
operator $\oj$ contains $J$ scalar fields, there are $J$ external
vertices on the Wilson loop, and the average distance between them
is of order $l/J$. A distance between the end-points of the
propagator $G(s,r)$ is, on average, of the same order: $|r-s|\sim
l/J$. The  expansion of the propagator  $G(s,r)$ in $(r-s)$
therefore generates $1/J$ expansion of the Feynman integral.

\section{String side}
\label{str}

In this section, we will develop a systematic method to solve equations
of motion in the sigma-model order by order in $1/j^2$ and to
compute the classical string action as a series in this parameter.

We can choose the coordinates $\tau$ and $s$ on the world sheet, such
that $s$ coincides with the natural parameter on $C$ when restricted to the
boundary at $\tau=0$.  We
can also assume that the world sheet has a topology of the cylinder,
and that the vertex operator is placed at $\tau=\infty$. The
action then takes the following
form\footnote{The other $S^5$ angle $\varPsi$ in \rf{smaction}
is zero on the classical
solution.}:
\ar{\label{ac1}
S &=& \f {1} {4 \pi j} \int_{\varepsilon}^{\infty} d \tau  \int_{0}^{2\pi l} ds\,
\left[\dt{P}{}^2 + \ds{P}{}^2+
\e^{2P}\br{\dt{X}{}^2+\ds{X}{}^2}
+ \dt{\varPhi}{}^2
+ \ds{\varPhi}{}^2  \right]
-\frac{l}{j\ep}
\nonumber \\* &&
\vphantom{S = \f {1} {4 \pi} \int_{0}^{\infty} d \tau  \int_{0}^{2\pi l} ds\,
\left[\dt{P}{}^2 + \ds{P}{}^2+ e^{2P}\br{\dt{X}{}^2+\ds{X}{}^2}
+ \dt{\varPhi}{}^2 + \ds{\varPhi}{}^2
\right]}
+\left . \lb P - i\varPhi \rb \right|_{\tau=\infty}\,
}
The cutoff on the integral over $\tau$ is necessary to regularize
the divergence of the area at the boundary of AdS space. The area
diverges as $2\pi l/\ep$. With our normalization of the action, we
should subtract $l/j\ep$ to make the action finite. As was shown
in \cite{MAL,DGO}, the simple subtraction is the correct way to
deal with the divergences in the AdS sigma-model, at least within
the semiclassical approximation.

The equations for the minimal surface should be supplemented by
boundary conditions. The boundary conditions at $\tau=0$ are set
by the Wilson loop:
\eq{
 X^{\mu}(s,0) = x^{\mu}(s),~~~~~ P(s,0) = + \infty,~~~~~\varPhi(s,0)=0.
} The source term in the action determines the behavior of the
string coordinates at infinity. Let $z$ be a local coordinate near
$\sigma_0$, so that the operator is inserted at $z=0$. Then $X^\mu$
are regular at $z=0$, but $P$ and $\varPhi$ have logarithmic singularities
determined by the source.
With the action normalized as in \rf{smaction}, we have: 
\eq{
P(z)=-j\ln|z|+{\rm regular},~~~~~\varPhi(z)=ij\ln|z|+{\rm regular}.
}
The exponential parameterization $z=\e^{(\tau + is)/l}$ 
maps the world sheet to a cylinder. The point of operator
insertion $z=0$ is mapped to infinity with the following boundary conditions
on the string coordinates:
\eq{
 X^{\mu}(s,\tau) \rightarrow x^{\mu}_{\infty},~~~~~
P(s,\tau) \rightarrow -\frac{ j\tau}{l},~~~~~\varPhi(s,\tau)\rightarrow
\frac{ij\tau}{l}~~~~~~~(\tau\rightarrow\infty).
}

We will solve the equations of motion perturbatively in $1/j^2$.
In order to do that, it is convenient to introduce new variables
in which the solution is non-singular at zeroth order (at infinite $j$).
We rescale:
\bqa
t = j \tau, \\
Q = P - \ln j. \label{whatisQ}
\eqa
In this way, we get:
\ar{\label{12}
 S &=&  \f 1 {4 \pi} \int_{j\ep}^{\infty} d t \, \int_{0}^{2\pi l} ds\,
\left[\dt{Q}{}^2 +\frac{1}{j^2}\, \ds{Q}{}^2+ e^{2Q}\br{j^2\dt{X}{}^2+\ds{X}{}^2}
+ \dt{\varPhi}{}^2 + \frac{1}{j^2}\,\ds{\varPhi}{}^2
\right]
\non
&&\vphantom{S =  \f 1 {4 \pi} \int_{0}^{\infty} d t \, \int_{0}^{2\pi l} ds\,
\left[\dt{P}{}^2 +\frac{1}{j^2}\, \ds{P}{}^2+ e^{2P}\br{j^2\dt{X}{}^2+\ds{X}{}^2}
+ \dt{\varPhi}{}^2 + \frac{1}{j^2}\,\ds{\varPhi}{}^2
\right]
}
-\frac{l}{j\ep}+\left . \lb Q - i\varPhi \rb \right|_{t=\infty}+\ln j\,.
}
The dynamics of the angular coordinate $\varPhi$ is trivial:
\beq
\varPhi = \frac{i t}{l}\,.
\eeq
Upon substitution of the solution for $\varPhi$,
the action combined with the vertex operator
(which just imposes the boundary condition for $Q$ at $t \to \infty$)
can be written in the following form:
\beq
\label{spr}
 S =  \f 1 {4 \pi} \int_{j\ep}^{\infty} d t \, \int_{0}^{2\pi l} ds\,
 \left[ \br{\dt{Q} + \f 1 l }^2 + \f 1 {j^2}\, \ds{Q}{}^2+ \e^{2Q} \left (j^2 \dt{X}^2+\ds{X}{}^2\right)
\right]+Q(j\ep)-\frac{l}{j\ep}+\ln j.
\eeq
Here, we used the equality
$$
\left.\br{Q(t)-i\varPhi(t)}\right|_{t=\infty}
=\frac{1}{4\pi} \int_{j\ep}^{\infty} d t \, \int_{0}^{2\pi l} ds\,
\frac{2}{l}\br{\dt{Q}+\frac{1}{l}}+Q(j\ep).
$$
Note that potentially dangerous boundary terms in \rf{12} have
been completely absorbed into the bulk part of the action. This
would have been impossible without cancellations between $AdS_5$
and $S^5$ contributions which occur because the vertex operator is
marginal. Potential divergences at $t\rightarrow\infty$ are of UV
nature, since they come from the vicinity of the operator
insertion, though in the coordinates we use they might look as an
IR effect. Anyway, they cancel and the action is saturated by
$t\sim 1$. The typical AdS scale $\e^{-P}$ is also small: $Q$
is finite at large $j$
 and $P\sim \ln j$ according to \rf{whatisQ}.  Hence, the largest contribution
to the action comes from the region of AdS space close to the boundary.

The equations of motions are
\bqa
\label{eom1}
-\dtt{Q}-\f 1 {j^2}\, \dss{Q} + \e^{2Q}\br{j^2\dtt{X}{}^2 +\ds{X}{}^2} = 0, \\
\label{eom2}
j^2\br{\e^{2Q} \dt{X}}\dt{\vphantom{X}}+\ds{\br{\e^{2Q}\ds{X}}}=0.
\eqa
They are readily solved at large $j$. First, we notice that
$X^\mu$ must be $t$-independent to the leading order. Hence, the minimal
surface is a cylinder with the contour $C$ as the base:
\eq{
\on{X^\mu}{0}(s,t)=x^\mu(s).
}
The equation \rf{eom1} reduces to
\eq{
-{\on{\dtt{Q}}{0}}+\e^{2\on{Q}{0}}=0,
}
which is solved by
\eq{
\on{Q}{0}=-\ln \left( l \sinh \f t l \right).
}

To develop a systematic procedure to solve the equations of motion
order by order in $1/j^2$, we write: \eq{ Q=\sum_{n=0}^{\infty} \f
1 {j^{2n}}\, \on{Q}{n},~~~~~ X^\mu=\sum_{n=0}^{\infty} \f 1
{j^{2n}}\, \on{X^\mu}{n}. } The structure of equations
(\ref{eom1}), (\ref{eom2}) is such that we can easily solve them
recursively. On each step we will need to solve an ordinary
linear differential equation for $\on{Q}{n}$, $\on{X^\mu}{n}$ with
an unknown dependence on time only. This is somewhat surprising,
since we are dealing with partial differential equations. The
reason for the simplification is an enhancement of time
derivatives by a factor of $j^2$ compared to the derivatives in $s$. 
The iterative solution can be
constructed as follows: suppose that we know $\on{Q}{m}$,
$\on{X^\mu}{m}$ for $m<n$. They must solve the first equation
(\ref{eom1}) with an accuracy $O(1/j^{2n})$, and the second
equation (\ref{eom2}) with an accuracy $O(1/j^{2n-2})$. The next
order in the second equation (\ref{eom2})  allows us to express a
linear combination of time derivatives of $\on{X^\mu}{n}$ through
the known functions. $\on{Q}{n}$ drops out at this order because
$\on{\dt{X}{}^\mu}{0}=0$. The first equation $(\ref{eom1})$ at
order $ 1/ {j^{2n}}$ then reduces to a
 linear equation for
$\on{Q}{n}$ and its second time derivative whose coefficients depend on
$\on{X^\mu}{n}$, which we already know, and on
$\on{Q}{m}$, $\on{X^\mu}{m}$ with $m<n$.

The iterative procedure is best exemplified by the first three steps:
\ar{
\label{hier}
\br{\e^{2\on{Q}{0}} \on{\dt{X}^\mu}{1}}\dt{\vphantom{X}}
&=&- \br{\e^{2\on{Q}{0}} \ds{\on{X^\mu}{0}{}}}\ds{\vphantom{X}} \Longrightarrow
\on{X}{1}^\mu, \non
\on{\dtt{Q}}{1} - 2 \rr \on{Q}{1}  &=& \rr
\br{ \on{\dt{X}}{1}^2 + 2 \ds{\on{X}{0}}\cdot\ds{\on{X}{1}}  } \Longrightarrow \on{Q}{1},
\non
\left[\rr\br{2\on{Q}{1} \on{\dt{X}^\mu}{1} +\on{\dt{X}^\mu}{2}}\right]\dt{\vphantom{X}}
 &=& -\rr\br{2\on{Q}{1} \ds{\on{X^\mu}{0}{}} + \ds{\on{X^\mu}{1}{}}}\ds{\vphantom{X}}
  \Longrightarrow  \on{X}{2}^\mu,
\\*
&\cdots& \nonumber
}
It is straightforward to integrate these equations, though calculations become
increasingly difficult with the order of iteration.
After rather lengthy algebra
we obtain:
\ar{
\on{X^\mu}{1}&=&\f {l^2}  4 \, \lb {\e^{-2\,\ti{t}}}+ 2 \,\ti{t}- 1 \rb\dss{{x^\mu}{}},
\non
\on{Q}{1}&=&-\f {l^2} 4\,{\frac {4\,\ti{t}{\e^{-2\,\ti{t}}}+{\e^{-4\,\ti{t}}}-1}{1-{\e^{-2\,\ti{t}}}}}
\,\dss{x}{}^2,
\non
\on{X^\mu}{2}&=&\frac{l^4}{16}\left[
\lb 2\,{\ti{t}}^{2}+2\,\ti{t}\,{\e^{-2\,\ti{t}}}+\,{\e^{-2\,\ti{t}}}-1 \rb\dsf{x^\mu{}}
\right.
\non
&&\left.
+4\lb \ti{t}\,{\e^{-2\,\ti{t}}}+ \,\ti{t}+\,{\e^{-2\,\ti{t}}}-1 \rb
\br{\dss{x}{}^2\ds{x^\mu{}}}\ds{\vphantom{x}}
\right.
\non
&&\left.
- \lb \,{\e^{-4\,\ti{t}}} + 8\,\ti{t}\,{\e^{-2\,\ti{t}}} + 4\,{\e^{-2\,\ti{t}}} + 4\,\ti{t}-5 \rb
\dss{x}{}^2\dss{x^\mu{}}
\right],
}
where $\ti{t}=t/l$.

The classical action can be written as a power series in
$1/j^2$:
\eq{\label{exps}
S=\ln j+\sum_{n=0}^{\infty}\frac{\on{S}{n}}{j^{2n}}\,
}
with\footnote{We can forget about regularization everywhere
except in $\on{S}{0}$.}
\ar{
\on{S}{0}&=&\frac{1}{4\pi}\int_0^{2\pi l}ds\,\int_{j\ep}^\infty dt\,
 \left[ \br{\on{\dt{Q}}{0} + \f 1 l }^2 +\rr \ds{\on{X}{0}}{}^2
\right]+\on{Q}{0}(j\ep)-\frac{l}{j\ep}\,,
\non
\on{S}{1}&=&\frac{1}{4\pi}\int_0^{2\pi l}ds\,\int_{0}^\infty dt\,
\rr\br{\on{\dt{X}}{1}^2+2\ds{\on{X}{0}}\cdot\ds{\on{X}{1}}},
\non
\on{S}{2}&=&\frac{1}{4\pi}\int_0^{2\pi l}ds\,\int_{0}^\infty dt\,
\rr\br{-\on{Q}{1}\on{\dt{X}}{1}^2+\on{\dt{X}}{1}\cdot\on{\dt{X}}{2}
+2\ds{\on{X}{0}}\cdot\ds{\on{X}{2}}}.
}
Calculating the integrals over $t$, we get: \ar{
\on{S}{0}&=&\ln\frac{2}{l}-1, \nonumber \\*
\on{S}{1}&=&-\frac{l}{8\pi}\int_0^{2\pi l}ds\, \dss{x}{}^2,
\nonumber \\* \on{S}{2}&=&\frac{l^3}{64\pi}\int_0^{2\pi l}ds\,
\left[2\dst{x}{}^2-\br{\dss{x}{}^2}^2\right]. } The appearance of
the dimensionful quantity under the logarithm in $\on{S}{0}$ may
seem strange, but in fact is required to reproduce  the correct
scaling dimension of the correlator, for which we get \ar{
\cor&\simeq&\frac{1}{J!}\,\br{\frac{\sqrt{\lambda}\,l}{2|x|^2}}^J
\exp \int_0^{2\pi l}ds\, \left\{\frac{\lambda l}{8\pi
J}\,\dss{x}{}^2 -\frac{\lambda^2l^3}{64\pi
J^3}\left[2\dst{x}{}^2-\br{\dss{x}{}^2}^2\right] \right. \non
&&\left. +O\br{\frac{\lambda^3}{J^5}} \right\}. } Here we used the
Stierling formula to approximate $\e^J/J^J\simeq 1/J!$. Expanding
the correlator in $\lambda$, we find the complete agreement with
the one-loop perturbation theory \rf{one-loop}! We also get a
prediction for the two-loop contribution. The two-loop calculation
on the SYM side is beyond the scope of the present paper. Instead,
we will check that a certain set of diagrams of all orders in
$\lambda$ exponentiate, as required by the general structure of
the correlator in string theory.

\section{Exponentiation}
\label{exp}

If we believe that the string description is valid at weak
coupling for large $J$, perturbative series for $\la W(C) O_J \ra$ must exponentiate.
This is a rather non-trivial statement, since the usual expansion in Feynman diagrams
computes the correlator itself, not its logarithm. Perturbative series
exponentiate for Abelian Wilson loops, but in the non-Abelian theory, especially
at large $N$, exponentiation is not at all obvious.
The string-theory prediction for the sum of all planar diagrams is
\beq \label{expon}
\la W(C) O_J (x)\ra\simeq \frac{1}{|x|^{2J}}
\,\e^{-J S(j;C)},
\eeq
where $S(j;C)$ has a regular expansion in $1/j^2=\lambda/J^2$, eq.~\rf{exps},
which can be regarded as the weak-coupling expansion.
 If we reexpand the correlator
in $\lambda$:
\eq{
\cor\simeq \frac{1}{J!}\left(\frac{\sqrt{\lambda}\,l}{2 |x|^2}\right)^J
\sum_{n=0}^\infty a_n\lambda^n,
}
we find that the contribution which is  least suppressed in $1/J$ at any given order
of perturbation theory is determined by just one number, $S_1$:
\eq{
a_n=\frac{S_1^n}{J^{n}n!}+O\br{\frac{1}{J^{n+1}}}.
}
This is because the exponent in \rf{expon} contains an overall factor of $J$,
and if we take the limit of large $J$ at a fixed order of perturbation theory,
only the contribution of the leading term in the expansion of $S(j)$
survives. Our goal will be to prove this universality by a direct
expectation of planar Feynman diagrams in the SYM theory.
To do that, we will consider diagrams with largest combinatorial factors
at a given order of perturbation theory.

\begin{figure}[h]
\begin{center}
\epsfxsize=3cm
\epsfbox{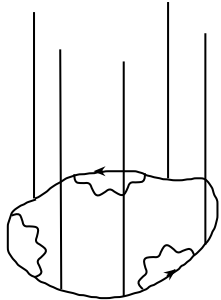}
\end{center}
\caption[x]{\small The diagrams with largest combinatorial
factors at large $J$.}
\label{pert2}
\end{figure}

The leading-order diagram contains $J$ scalar propagators that
connect the local operator to $J$ points on the Wilson loop. In view
of the standard equivalence of planar Feynman graphs and
discretized two-dimensional surfaces \cite{'tHooft:1973jz},
this diagram can be thought of as a prism with $J$ facets. The base of the
prism is the Wilson loop and its apex is the point where
 operator $\oj$ is inserted. This point is moved to infinity
in our approximation. Higher-order diagrams
correspond to decorating
the facets with extra propagators and vertices. Until the order of
perturbation theory becomes comparable to $J$,
most of the facets will remain empty.
It is clear that adding a new element to one of the empty facets produces
a combinatorial factor of order $J$, while adding a propagator or a vertex to
an already decorated facet produces
a combinatorial factor of order one. Hence, diagrams with the biggest combinatorial
factor have $n$ propagators distributed among $n$ different facets.
Diagrams of this type with internal interaction vertices cancel by
a trivial extension of the argument in appendix~B. These cancellations
leave only the diagrams with $n$ loop-to-loop propagators sandwiched between
$J$ external legs (fig.~\ref{pert2}).

Thus, to the leading order in $1/J$:
\ar{
a_n&=&\frac{(J-1)!}{(2\pi l)^J}\int dq_1\ldots dq_J\, \theta_c(q_1,\ldots,q_J)
\non
&&\times
\sum_{1\leq p_1\leq\ldots\leq p_n\leq J}
\int_{q_{p_1}}^{q_{p_1+1}} ds_1\int_{s_1}^{q_{p_1+1}} dr_1
\ldots
\int_{q_{p_n}}^{q_{p_n+1}} ds_n\int_{s_n}^{q_{p_n+1}} dr_n\,
\non
&&\times
G(s_1,r_1)\ldots G(s_n,r_n)
}
We can change the order of integration and integrate
over $q_i$ using eq.~\rf{s!!}. Then:
\ar{
a_n&=&\frac{(J-1)!}{(2\pi l)^J}\int ds_1 dr_1\ldots ds_n dr_n\,\theta_c(s_1,r_1,\ldots,s_n,r_n)
\,G(s_1,r_1)\ldots G(s_n,r_n)
\non
&&\times
\frac{J}{n}
\sum_{k_1+\ldots+k_n=J}\frac{(s_2-r_1)^{k_1}\ldots(s_1-r_n)^{k_n}}{k_1!\ldots k_n!}
\non
&=&\frac{1}{n\,(2\pi l)^J}\int ds_1 dr_1\ldots ds_n dr_n\,\theta_c(s_1,r_1,\ldots,s_n,r_n)
\,G(s_1,r_1)\ldots G(s_n,r_n)
\non
&&\times
\left[2\pi l-(r_1-s_1)-\ldots-(r_n-s_n)\right]^J.
}
Applying \rf{f0} repeatedly $n$ times and keeping only the leading order
in $1/J$, we get
\ar{
a_n&\approx&\frac{(2\pi l)^{n}}{n(J+1)\ldots (J+n)}
\int ds_1 \ldots ds_n \,\theta_c(s_1,\ldots,s_n)
\frac{\dss{x}{}^2(s_1)}{16\pi^2}\ldots\frac{\dss{x}{}^2(s_n)}{16\pi^2}
\non
&\approx&\frac{1}{n!}\br{\frac{l}{8\pi J}\int ds\,\dss{x}{}^2}^n,
}
in agreement with the prediction of string theory: the diagrams with largest
combinatorial weights indeed exponentiate.

\section{Discussion}\label{last}

We made a rather detailed comparison between diagrams of perturbation
theory and the semiclassical string amplitude which both compute
a two-point correlator of an arbitrary Wilson loop with a local operator
in \N SYM theory. The results
completely agree to the one-loop accuracy, as well as to all orders in perturbation theory
for  diagrams with the largest combinatorial weights.
We should stress once again that there are no apparent reasons for
such an agreement. The string theory and the perturbation theory
compute different regimes, which
can be most easily seen from the general form of the semiclassical
string amplitude \rf{gstring}. On the string side, we have to assume that
the exponent in  \rf{gstring} is large, otherwise the semiclassical approximation
breaks down. To compare with SYM perturbation
theory, we expand the exponential and thus assume that the exponent is
small. Perhaps, one can invoke arguments similar to those of
\cite{Tseytlin:2002ny}
to explain the agreement between the two calculations. The arguments rely
on the observation that the 't~Hooft coupling appears only in the combination
$\lambda/J^2$ on both sides of the correspondence, which for example
means that $\lambda/J^2$, and not $\lambda$ itself, is a parameter of the planar
perturbation theory. Still, a direct comparison to the string theory requires
a non-trivial resummation of perturbative series.

Incidentally, we found that the string calculations are technically
simpler than perturbative SYM calculations. It is relatively easy
to compute the string amplitude to the two-loop order
and it is definitely possible to push string calculation to higher
orders in $\lambda/J^2$, while on the SYM side already a two-loop calculation
constitutes an enormously hard problem. The simplicity of
the string calculation may indicate that various diagrams
cancel leaving a simple net result.
We indeed observed cancellations
at one loop, but those look
rather accidental, at least in the way we found them.

\subsection*{Acknowledgements}

We are grateful to A.~Tseytlin for discussions.
 One of us (V.P.) would like to thank the Department of Theoretical
Physics at
 Uppsala University, where this work was partially done,
for kind hospitality.
 This work was supported in part by STINT grant IG 2001-062.
 V.P. was also supported in part by
INTAS grant 00-561, RFBR grant 01-02-17488a,  RFBR grant
 for Support of Young Scientist 02-02-06517, and  Russian
 President's grant 00-15-99296.
K.Z. was supported in part by
 RFBR grant 01-01-00549, and grant
 00-15-96557 for the promotion of scientific schools.

\setcounter{section}{0}
\appendix{$N=4$ SYM theory}

In this appendix we summarize our notations and conventions
for the SYM perturbation theory.
Feynman rules follow from
the SYM action
\eq{\label{action}
S=\frac{N}{\lambda}\int d^4x\,\tr\left\{
\frac12\,F_{\mu\nu}^2+\br{D_\mu\Phi_i}^2-\frac12\,[\Phi_i,\Phi_j]^2
+{\rm fermions}
\right\}.
}
We do all calculations in the Feynman
gauge in the coordinate
representation, where the propagators are
\beq
\la \Phi_i^{AB}(x_1) \Phi_j^{CD} (x_2) \ra =
\f {g^2} {8 \pi^2} \f {\vd^{AC} \vd^{BD} \vd_{ij}} {|x_1-x_2|^2},
\eeq
\beq
\la A_{\mu}^{AB}(x_1) A_{\nu}^{CD} (x_2) \ra =
\f {g^2} {8 \pi^2} \f {\vd^{AC} \vd^{BD} \vd_{\mu \nu}} {|x_1-x_2|^2}.
\eeq
The capital letters denote $U(N)$ indices.

\appendix{Cancellation of diagrams with internal vertices}
\label{intvert}

\begin{figure}[h]
\begin{center}
\epsfxsize=13cm
\epsfbox{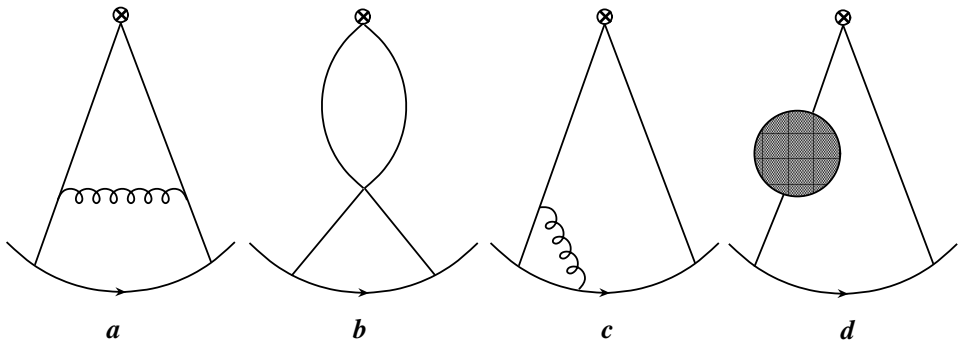}
\end{center}
\caption[x]{\small One-loop diagrams with internal vertices.}
\label{dmy}
\end{figure}

There are four types of diagrams (fig.~\ref{dmy}) that were not
taken into account in sections~\ref{SYMside} and \ref{exp}. We
show here that they cancel in the large-distance asymptotic of the
correlator $\vev{W(C)\oj(x)}$. We can amputate those legs of the
propagators in diagrams (a), (b), and (c) which couple to the
local operator and replace them by $1/(8\pi^2|x|^2)$. Indeed, the
region of integration
with intermediate
points close to $x$ corresponds to
renormalization of the operator $\oj$. Since the operator is chiral and
is not renormalized, these contributions mutually cancel. The
intermediate region of integration corresponds to descendants and
contributes at higher orders in $1/|x|$. Thus we amputate
the external legs and also take into account one half of the
diagram (d). Another half participates in cancelling the renormalization
of the operator $\oj$.

We start with the diagram (d).
The wave function renormalization
does not vanish in the Feynman gauge, and was computed
in \cite{Erickson:2000af}.
We can take the wave-function renormalization into account
by multiplying each propagator in the tree-level amplitude by
a factor $1+F_d$, where
\eq{
F_d=-\lambda\int d^4x\,D^2(x),
}
and $D(x)$ is appropriately regularized scalar propagator, which
goes to $1/(4\pi^2|x|^2)$ when regularization is removed.

The diagram (a) with external legs amputated
gives a factor of
\eq{
F_a(x_1,x_2)=\frac{\lambda}{2}\int d^4y\,d^4z\,\,D(y-z)\,\frac{\d}{\d y^\mu}D(y-x_1)\,
\frac{\d}{\d z^\mu} D(z-x_2).
}
for each pair of external legs. Here, $x_1$ and $x_2$ are
adjacent vertices on the contour $C$.
The diagram (b) gives
\eq{
F_b(x_1,x_2)=\frac{\lambda}{2}\int d^4y\,D(y-x_1)D(y-x_2).
}
Finally, the diagram (c) contributes
\eq{
F_c(x_1,x_2)=\frac{\lambda}{2}\int_{s_1}^{s_2} ds\,\ds{x}^\mu(s)\int d^4y\,
D(y-x)\,\frac{\d}{\d y^\mu}\Bigl(D(y-x_2)-D(y-x_1)\Bigr),
}
where $x(s)$ is a point on the contour between $x_1\equiv x(s_1)$ and
$x_2\equiv x(s_2)$. Integration by parts
yields:
\eq{
F_c=-F_d-2F_b.
}
This equation, together with  $F_a=F_b$, implies
that
\eq{
F_a+F_b+F_c+F_d=0.
}

\end{document}